\documentclass{article}
\usepackage{spconf,amsmath,graphicx}
\usepackage{caption,subcaption}
\usepackage{cite}
\usepackage{hyperref}

\title{Acoustically-Driven Phoneme Removal That \\Preserves Vocal Affect Cues}
\twoauthors
 {Camille Noufi$^*$\thanks{Author correspondence: \url{cnoufi@ccrma.stanford.edu}}, Jonathan Berger}
	{Stanford University\\
	Center for Computer Research \\in Music and Acoustics\\
	Stanford, CA, USA}
 {Karen J. Parker, Daniel L. Bowling\sthanks{This work was funded in part thanks to a grant from the National Institute of Mental Health (K01MH122730) and \href{https://neuroscience.stanford.edu/research/funded-research/quantifying-auditory-vocal-affect-human-social-communication}{a seed grant from the Wu Tsai Neurosciences Institute at Stanford University.}}}
	{Stanford School of Medicine\\
	Department of Psychiatry \\and Behavioral Sciences\\
	Stanford, CA, USA}
\begin{document}
\ninept
\topmargin=0mm
\maketitle
\begin{abstract}
In this paper, we propose a method for removing linguistic information from speech for the purpose of isolating paralinguistic indicators of affect. 
The immediate utility of this method lies in clinical tests of sensitivity to vocal affect that are not confounded by language, which is impaired in a variety of clinical populations. 
The method is based on simultaneous recordings of speech audio and electroglottographic (EGG) signals. 
The speech audio signal is used to estimate the average vocal tract filter response and amplitude envelop. 
The EGG signal supplies a direct correlate of voice source activity that is mostly independent of phonetic articulation. 
The dynamic energy of the speech audio and the average vocal tract filter are applied to the EGG signal create a third signal designed to capture as much paralinguistic information from the vocal production system as possible---maximizing the retention of bioacoustic cues to affect---while eliminating phonetic cues to verbal meaning. 
To evaluate the success of this method, we studied the perception of corresponding speech audio and transformed EGG signals in an affect rating experiment with online listeners. 
The results show a high degree of similarity in the perceived affect of matched signals, indicating that our method is effective.
\end{abstract}
\begin{keywords}
speech, paralanguage, affect, voice transformation, electroglottagraphy, phoneme removal
\end{keywords}
\section{Introduction}
\label{sec:intro}
Much of the information conveyed by speech is transmitted through paralinguistic cues encoded in the audio signal. 
These paralinguistic cues are essential to the communication of emotions, intentions, and personality ~\cite{Kreiman2011FoundationsPerception,Schuller2014ComputationalProcessing,VanZant2020HowPersuades}. 
For the majority of our daily interactions, these paralinguistic cues are embedded among phonetic cues encoding linguistic meaning. 
Although most individuals have no problem parsing linguistic and paralinguistic cues in speech and responding appropriately, this ability is often impaired in clinical populations (e.g., in autism~\cite{OConnor2012AuditoryReview} and depression~\cite{Naranjo2011MajorStimuli}). 
Focusing on autism, the impairment is assumed to pertain to the reception of paralinguistic cues to speaker affect. 
However, the tests on which this assumption is based use speech stimuli, and thus confound sensitivity to paralanguage with language functioning. 
Testing sensitivity to paralinguistic affect directly requires isolating it from speech. 
This is important for understanding the nature of auditory-vocal contributions to clinical dysfunction, particularly in mental health.

Existing methods that attempt to isolate paralinguistic cues from speech benefit from economy and efficiency, but they also lose significant amounts of paralinguistic information, particularly concerning affect. 
For example, one simple and efficient method is to remove phonologic content from speech audio by adaptively low-pass filtering the signal such that the filter roll-off occurs below the second formant peak, thus removing a critical cue to vowel identification (i.e., the ratio between the first and second formants). 
However, because this method removes high-frequency content ($>$ starting at approx. 500-2500 Hz, depending on the vowel~\cite{Ladefoged2019ElementsPhonetics}), it also destroys important affective content~\cite{Guzman2013InfluenceExpression}.
\begin{figure*}[th!]
    \centering
    \includegraphics[width=0.95\textwidth]{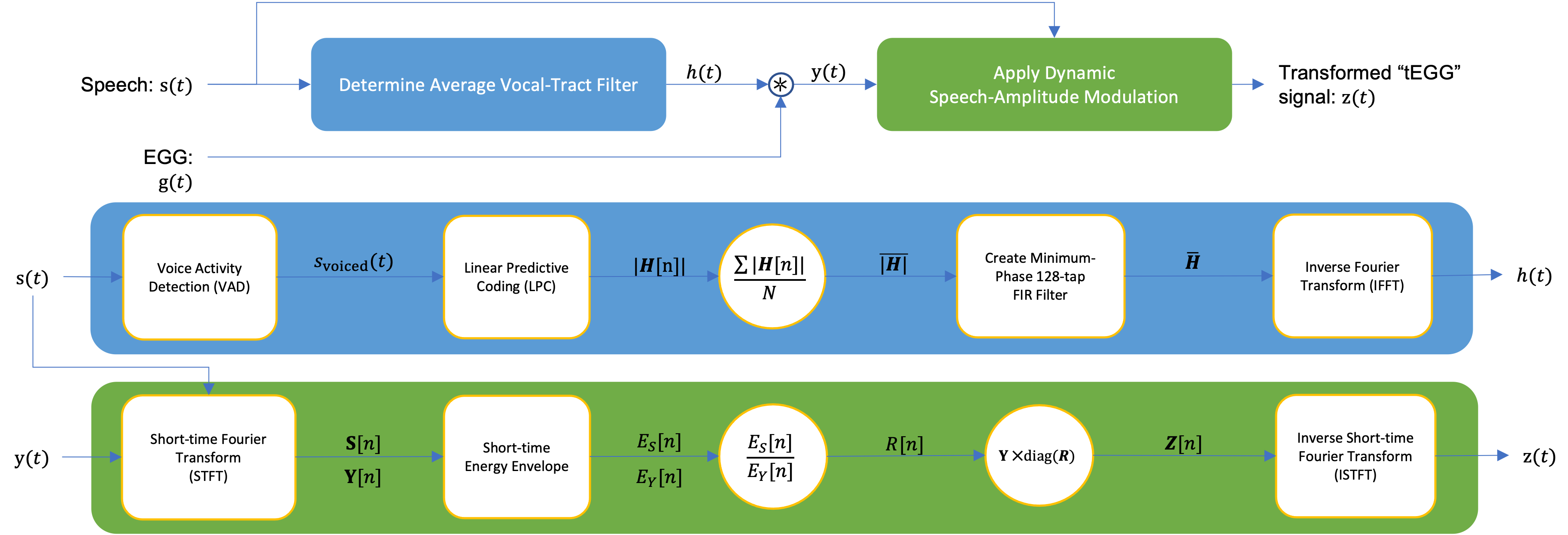}
      \caption{Proposed transformation method given speech signal $s(t)$ and EGG source signal $g(t)$. The two-stage method creates signal $z(t)$ that preserves paralinguistic content contained in the source, average vocal tract shape, and amplitude envelope, while removing linguistic content supplied by supralaryngeal articulation. The first stage (blue) extracts the LPC-based average resonant impulse response $h(t)$ representing the vocal tract during periods of vocal source activity. $h(t)$ is convolved ($*$) with $g(t)$ in the second stage (green) and dynamically scaled via cross-filtering to produce $z(t)$.}
    \label{fig:flowchart}
\end{figure*}

Another method is to discard phonetic cues by separating the vocal signal into two parts, the signal representing the laryngeal source, and the signal representing the supralaryngeal filter. 
Whereas the filter is more typically associated with linguistic articulation~\cite{Hill1977APhonetics,Gafos2020Editorial:Production,Zhang2016MechanicsControl}, the source is more associated with paralinguistic features that are essential to affect, such as voice pitch, breathiness, roughness, and other varieties of voice quality~\cite{Kreiman2011FoundationsPerception, Kreiman2021, Scherer1978PersonalityExtroversion,Anikin2020AVocalizations, VanZant2020HowPersuades}. 
The most common method for separating the vocal source signal $e(t)$ from the vocal tract impulse response $h(t)$ is linear predictive coding (LPC)~\cite{Shaughnessy1988LinearCoding}. 
LPC uses a $p^{th}$-order linear predictor to estimate speech signal $\Tilde{s}(t)$ from $p$ previous samples. 
This is done by solving for optimum predictor coefficients $a_k$ of the filter for a pseudo-stationary frame of speech. 
Once the filter coefficients are found the “residual” $e(t)$ is calculated, representing a mixture of both glottal and noise-based phonetic content.

Although source-filter separation is useful in many cases, the reality is that both linguistic and paralinguistic information are encoded across the entire range of the frequency spectrum~\cite{Guzman2013InfluenceExpression} and produced by both source and filter~\cite{Scherer1978PersonalityExtroversion,Kreiman2011FoundationsPerception,Kreiman2021,Anikin2020AVocalizations}. 
Given that existing methods operating on speech audio alone do not adequately separate linguistic and paralinguistic information, we designed a new method that leverages synchronized audio and electroglottagraphic (EGG)~\cite{Lecluse1975TheActivity,Childers1984RelationshipsContact,Hong1997ElectroglottographySpeech,Herbst2020ElectroglottographyUpdate} recordings of speech to create a third transformed EGG (“tEGG”) signal made by applying a speech-based spectro-temporal transform to the EGG signal. 
When played as audio, this stimulus lacks the speech signal’s linguistic content but retains significant paralinguistic information across the frequency spectrum. 
In Section \ref{sec:method}, we describe the algorithm for the transformation method. 
In Section \ref{sec:implementation}, we describe the speech/EGG recording process and an implementation of the algorithm. 
In Section \ref{sec:evaluation}, we evaluate our algorithm in an online experiment comparing ratings of emotional affect in corresponding speech and tEGG signal pairs. 
We conclude by briefly summarizing future directions and applications of the method introduced here for understanding and computing vocal affect. 

\section{Transformation Method}\label{sec:method}
The transformation algorithm operates in two stages (Figure \ref{fig:flowchart}). 
The first stage, represented by the blue block in Figure \ref{fig:flowchart}, aims to preserve the average frequency response profile of the speaker’s vocal tract during voice speech production, up to, but not including, the supralaryngeal articulation. 
This involves iteratively estimating the filter representation of the vocal tract through LPC on short frames of audio, taking an average of the results, and then convolving this average filter representation in the time domain with the source signal captured by the EGG. 
The second stage, represented by the green block in Figure \ref{fig:flowchart}, aims to preserve the dynamic energy envelope of the original speech audio signal. 
This involves extracting dynamic amplitude modulation over time from the speech signal and applying it to the convolved signal.

Let a speech audio signal $s(t)$ and a corresponding EGG signal $g(t)$ both be monophonic digital audio signals with sampling rate $f_s$. 
Stage one of the algorithm proceeds as follows. An energy-based voice activity detection algorithm~\cite{Brookes2011} is used to indicate if a sample at time $t$ contains voice activity or not. 
Samples containing voice activity are concatenated and the remaining samples are discarded.
We perform LPC on short frames of audio (assumed to be pseudo-stationary) to obtain the per-frame filter coefficients $A_v$, $A_g$, and $A_l$ representing the vocal tract, glottal source, and lip radiation filters, respectively.
This algorithm uses an Iterative Adaptive Inverse Filtering method based on a Glottal Flow Model to estimate the linear prediction coefficients of both vocal tract and glottis filters from a speech signal frame~\cite{Perrotin2019ASpeech}.
An adaptive filter with order $2 + f_s/1000$ is used to determine $A_v$ and a 3rd-order filter to determine $A_g$. 
We apply a Hamming window to each frame of the voiced speech audio signal to obtain the short frame $s_n(t)$.
To estimate the complex frequency response of the vocal tract, we set the the coefficients $A_v$ as zeros of the complex polynomial $H_n(e^{j\omega})$ and set the polynomial representation of the poles to 1.
For frames $1 ... N$, we apply LPC to each overlapping frame of voiced samples to obtain the vocal tract frequency response at each $n$th frame.
We subsequently calculate the average of the magnitudes of the complex frequency responses, $|\overline{H[n]}| = \frac{\sum_{n=1}^{N} |H_n(e^{j\omega})|}{N} $.
A minimum-phase finite impulse response (FIR) filter is created given $|\overline{H[n]}|$.  
The phase of the FIR filter is the imaginary component of the negative of Hilbert transform of the magnitude of the full spectrum frequency response.
Finally, the estimated average impulse response of the vocal tract $h(t)$ is convolved with the glottal source signal $g(t)$ in the time domain to create filtered source signal $y(t)$.

In stage two of the algorithm, we then dynamically modulate the amplitude of the convolved signal $y(t)$ to match the amplitude of the original speech signal $s(t)$. 
We apply a short-time Fourier transform (STFT) to obtain time-frequency representations $S[n]$ and $Y[n]$.  
We obtain the short-time energy envelopes $E_S[n]$ and $E_Y[n]$ of each representation by taking the sum of the magnitudes across the frequency spectrum at each time step $n$.  
We determine the dynamic modulation vector $R[n]$ to be the ratio of the short-time energy envelopes $E_S[n]$ and $E_Y[n]$.  
This dynamic modulation vector is then applied to $Y[n]$ via a right-hand-side multiplication to yield $Z[n]$.
Equations \ref{eq:cross-synth} and \ref{eq:ratio} describe this cross-filtering procedure:
\begin{equation}
    \textbf{Z} = \textbf{Y} \times \text{diag}(\textbf{R})
    \label{eq:cross-synth}
\end{equation}
where $\textbf{Z}^{F \times N}$ and $\textbf{Y}^{F \times N}$ are real-valued matrices and $\textbf{R}^{1 \times N}$ is a real-valued column vector.  $F$ represents the number of frequency bins $f$ determining the resolution of the STFT. Each $n$th element of $\textbf{R}$ is determined by   
\begin{equation}
     R[n] = \frac{E_S[n]}{E_Y[n]} = \frac{\sum_{f=1}^{F} |S_f[n]|}{\sum_{f=1}^{F} |Y_f[n]|}
     \label{eq:ratio}
\end{equation}
$Z[n]$ now represents the EGG source signal that has been both filtered by the average vocal tract response and the dynamic energy envelope of the speech signal.
The inverse STFT is applied to bring back $Z[n]$ into the time domain, resulting in the tEGG signal $z(t)$ in which phonetic variation has been systematically removed.

\section{Implementation}\label{sec:implementation}
To test the transformation method, we used a series of parallel audio and EGG recordings of affective speech made by our group as part of a related project~\cite{Bowling22_TAVA}. Here we provide only the details of the protocol that apply to implementing our transformation method. 

In brief, “voice actors” from the local Stanford community were recruited to produce a set of audio and EGG recordings of affective speech. Each actor was tasked with expressing 16 varieties of affect using a short emotionally neutral sentence (representative examples include “He stands on the dock”  and ``A bag is in the room”~\cite{Ben-David2011AInjury}). Speech recording sessions took place inside a sound-attenuating chamber at the Center for Computer Research in Music and Acoustics at Stanford University. After a short vocal warm-up and EGG calibration, the experimenter sounded a clapperboard (as a synchronization pulse) and proceeded to lead the actor through the 16 affect targets in a fixed order. Actors were recorded standing.

Audio recordings were made with a microphone (DPA, 466-OC-R-B00) and a digital audio recorder (Zoom H4n PRO) set to sample at 48 kHz with 24 bit-depth. The microphone was located 30 cm in front of the actor’s mouth. Parallel EGG recordings were made with an electroglottagraph (Glottal Enterprises, model EG2-PCX2) and a pair of skin surface electrodes placed on either side of the larynx and held in place by an elastic strap. The EGG signal was A-to-D converted by the electroglottagraph and transmitted over USB to a computer (Apple Macmini 9,1) running Adobe Audition (v 13.0.13.46) set to sample at 48 kHz at 4 bit depth. The sound intensity level was recorded in dB using a sound level meter (NTi XL2 with M4261 microphone; frequency weighting = A, sample rate = 10 Hz).

The audio, EGG, and sound intensity tracks from each recording were synchronized in post-processing. Then, for each actor, the experimenters listened and selected a single rendition basis of its perceived conveyance of the intended affect and being free of artifacts (e.g., coughing or failed electrode contact). Next, electrical noise was removed from each selected rendition’s EGG channel using Audition’s “Effects$\rightarrow$DeNoise” function and a proximate voice-free sample of the noise. EGG channels which contained substantial low-frequency energy below the fundamental frequency of the voice (due, e.g., to laryngeal movement) were additionally high-pass filtered with an adaptive cut-off set approximately 20 Hz below the fundamental. Finally, the amplitudes of the audio and EGG tracks were adjusted to accord with measured sound intensity level (at 60\% scale, to avoid clipping), and the mean (calculated across all recordings) was centered at -23 LUF using the EBU R-128 loudness standard ~\cite{Matlab,2020LoudnessSignals}. The results were set between two 50 ms buffers of silence and saved as the left and right channels of a two-track .wav file.

Applying the transformation method to these tracks first required creating a copy of the speech audio signal $s(t)$ down-sampled to 16 kHz. This was necessary because the adaptive filtering procedure used in LPC is not guaranteed to produce reliable results on high frequency content. All steps to compute the average vocal tract filter response via LPC were performed on this 16 kHz signal. We used a 20 ms Hamming window, 50\% overlap, and adjusted endpoints to adhere to constant-overlap-add requirements~\cite{Smith2011Spectral}. The resulting impulse response $h(t)$ was then upsampled to the original sampling rate of 48 kHz using 4th-order low-pass filter interpolation. This upsampled impulse response was then convolved with $g(t)$ and dynamic amplitude modulation was performed at 48 kHz. The two-stage transformation process runs approximately 30\% faster than real-time.

\begin{figure}[tb!]
     \centering
     \includegraphics[width=\columnwidth]{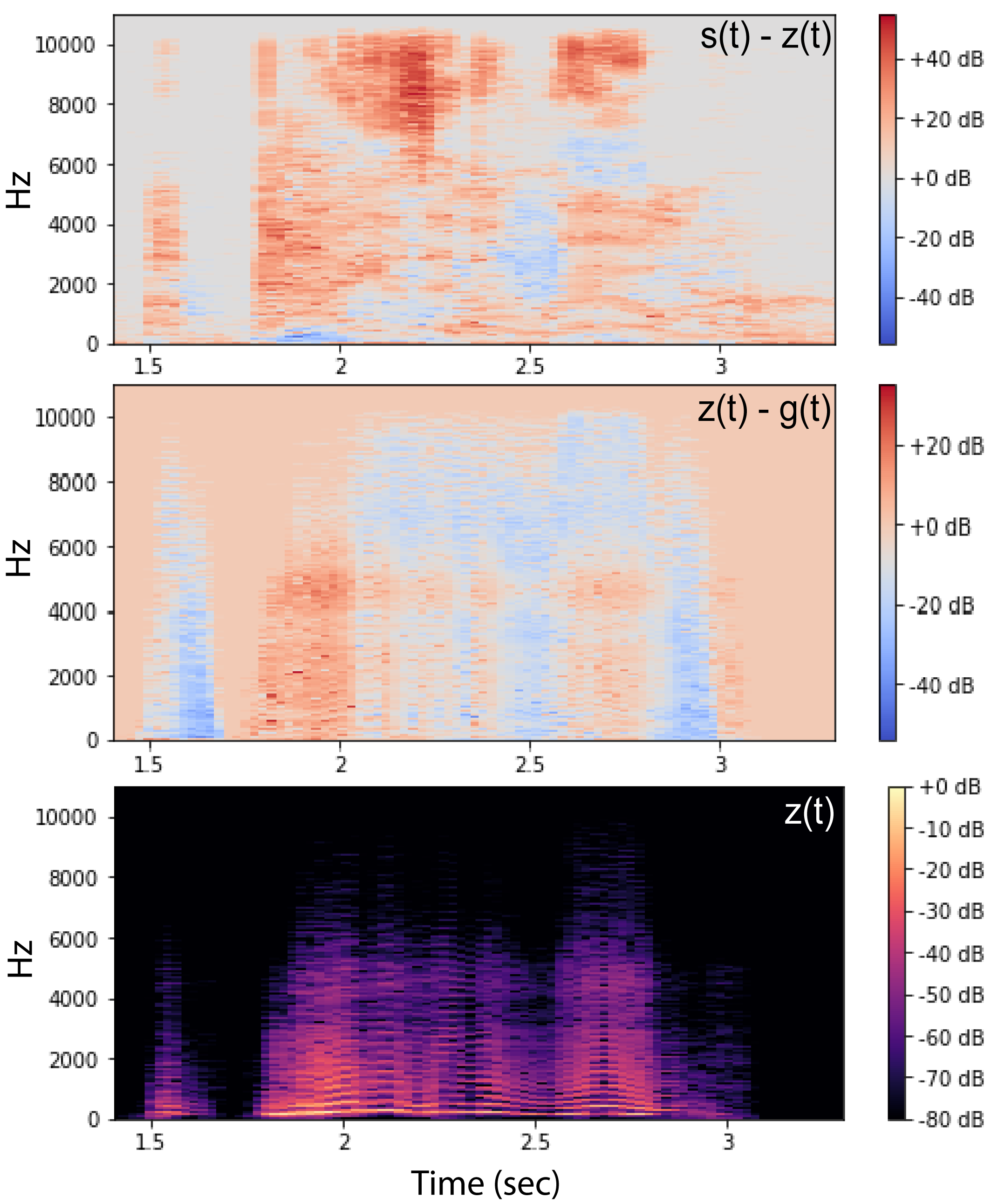}
     \caption{Residual difference spectrograms for affective speech sample $s(t)$. The sample uses “A bag is in the room” as linguistic material.  (Top) speech audio signal $s(t)-$ the tEGG signal $z(t)$; (Middle) $z(t)-$ EGG signal $g(t)$. (Bottom) transformed tEGG signal $z(t)$. These spectrograms detail the signal content removed and augmented through the transformation process. }
     \label{fig:spectrograms}
\end{figure}

The tEGG signal $z(t)$ is the resulting audio signal that retains the energy levels, rhythmic cadence, pitch prosody and voice quality of the speech signal at a 20 ms resolution. The residual spectrograms in Figure \ref{fig:spectrograms} show the acoustic content removed from $s(t)$ and added to $g(t)$ to obtain $z(t)$. As shown in Figure \ref{fig:spectrograms}, the average resonant filter $h(t)$ applied to the source signal $g(t)$ allows the signal to retain the speaker’s average ``tone of voice” as dictated by anatomy of the vocal tract, but removes dynamic variations in the formants that result in phonologic variation. Instead, the result is an average of all the vowels and voiced consonants uttered in the sentence.\footnote{\href{https://ccrma.stanford.edu/~cnoufi/demos/TAVA-PhonemeRemoval}{Audio examples of the transformation process, alongside comparative baselines} and \href{https://github.com/camillenoufi/TAVA-LanguagePerceptionRemoval}{code repository}, can be found online: \url{https://ccrma.stanford.edu/~cnoufi/demos/TAVA-PhonemeRemoval/}}

\section{Evaluation}\label{sec:evaluation}
To determine the effects of our transformation method on perception, we conducted an online affect rating experiment. We hypothesized that affect ratings for the transformed (tEGG) signals $z(t)$ will be practically the same as those for matching original speech audio signals $s(t)$. Statistically, we tested this by comparing variation in ratings within speech-tEGG signal pairs $[s(t), z(t)]$ to variation in ratings of the same speech audio signal. We reasoned that, if variation within signal pairs is $\leq$ variation in ratings of the speech audio signal, it would imply that our transformation method successfully preserves the major paralinguistic cues to affect in speech.

For this experiment, a total of 61 participants were recruited from Amazon’s Mechanical Turk, via Cloud Research (mean age=40, SD =12; 31 male, 28 female, 1 unknown). Each participant rated 128 stimuli (64 $[s(t), z(t)]$ pairs). These 64 pairs were derived from a set of 64 speech audio stimuli selected from a larger set on the basis of being consistently rated for affect across listeners in a different study~\cite{Bowling22_TAVA}. They included 21 different voices (10 male), with four 4 instances of each of the 16 targets.

\begin{figure}[tb!]
    \begin{subfigure}[b]{0.99\columnwidth}
     \centering
     \includegraphics[width=\columnwidth]{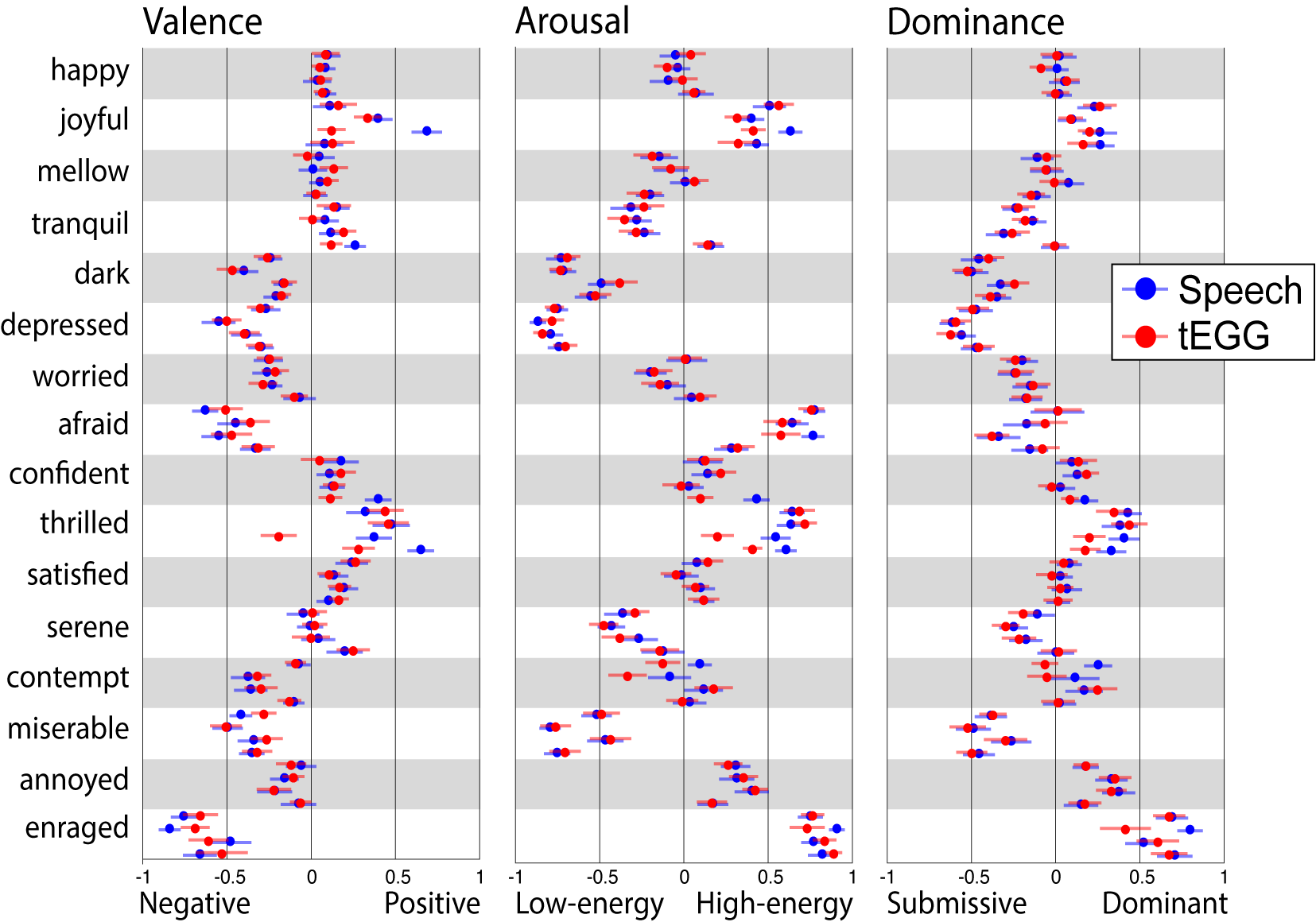}
     \caption{Mean ratings and 95\% confidence intervals for 64 speech-tEGG signal pairs.}
     \label{fig:ratings}
    \end{subfigure}
    \begin{subfigure}[b]{0.99\columnwidth}
         \centering
         \includegraphics[width=\columnwidth]{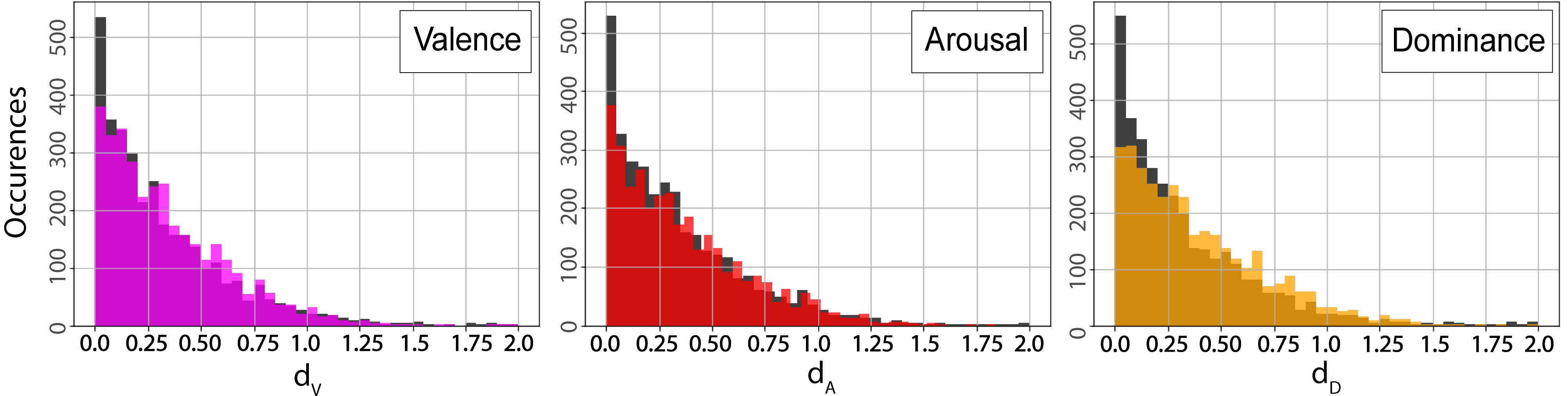}
         \caption{Comparison of rating difference scores to speech-only baseline. The dark gray histogram displays the distribution of difference scores calculated for speech-tEGG pairs, e.g. $d_V = |V_{s(t)}–V_{z(t)}|$ for valence.  The overlaid color histogram displays the distribution of difference scores for pairs of ratings (e.g. $d_{V_\text{baseline}} = |V1_{s(t)}–V2_{s(t)}|$ of the same speech signal $s(t)$).}
         \label{fig:histograms}
     \end{subfigure}
     \caption{Affect ratings of valence ($V$), arousal ($A$) and dominance ($D$) given to speech signals $s(t)$ and transformed signals $z(t)$ by listeners ($N=53$ raters).}
 \end{figure}

For each trial within an experiment, a single stimulus was rated using three visual analogue scales: valence (V; from ``negative” to ``positive”), arousal (A, from ``low-energy” to “high-energy”), and dominance (D; from ``submissive" to ``dominant”)~\cite{Laukka2005AEmotion}. Stimulus order was fully randomized and participants were not shown the 16 categorical affect labels. The experiment began with 10 practice trials (half speech, half tEGG, not paired) designed to familiarize participants with the rating interface. Practice stimuli did not overlap with the 128 test stimulus set, no feedback was given, and these data are not analyzed here. Finally, a subset of 10 stimuli from the 128 test stimulus set (half speech, half tEGG) were repeated throughout the experiment. This allowed us to assess intra-rater reliability as an indicator of whether or not participants performed the task in good faith. Data from 7 of the 61 participants were discarded on these grounds to poor correlation across repeated ratings (mean intra-rater correlation coefficient $< 0.5$). Thus, 20,352 ratings (53 raters $\times$ 128 stimuli $\times$ 3 dimensions) were used in subsequent analyses. The average affect ratings for each signal are shown in Figure \ref{fig:ratings}.

For every speech-tEGG stimulus pair, we calculate absolute differences in affect ratings along each affect dimension ($d_V$, $d_A$, and $d_D$), resulting in a set of 3,392 differences (64 stimulus pairs $\times$ 53 raters) for each dimension. As a basis for comparison with these data, we estimated ``baseline” variation in speech-affect perception by randomly selecting 53 pairs of ratings for each of the 64 speech signals, and calculating the differences along each dimension as above. This random sampling procedure was repeated 20 times to guard against possible sampling bias. Figure \ref{fig:histograms} shows the probability distribution of the observed speech-tEGG differences for each dimension (in gray) overlaid with a representative probability distribution of differences in affect ratings for the same speech stimuli (in color). A close correspondence between distributions in apparent in every case, as it was across all 20 repetitions.

As a basic evaluation of the statistical significance of these correspondences, we compared the speech-tEGG differences to the baseline difference using two-tailed Mann Whitney $U$-tests for each repetition~\cite{Scipy_MWU}. The results were not significant for valence ($U_\mu = 5.86\text{e}6, n_1=n_2=3392, P_\mu = 0.209$) or arousal ($U_\mu = 5.91\text{e}{6}, n_1=n_2=3392, P_\mu = 0.141$), indicating that variation in affect perception between paired speech and tEGG stimuli was statistically indistinguishable from baseline variation in speech-affect perception along these dimensions. For dominance, there was a significant difference ($U_\mu =5.93\text{e}6, n_1=n_2=3392, P_\mu < 1e-3$), but its direction was such that variation in affect perception between paired speech and tEGG stimuli was less than baseline variation in speech-affect perception (medians $= 0.23$ vs. 0.33, respectively). More telling than these tests is a measure of absolute effect size~\cite{Cohen2013StatisticalSciences}. For valence, Cohen’s $d$ varied from -0.065 to 0.009 $(\mu=-0.036)$ across the 20 repetitions; for arousal, it varied from -0.07 to -0.03 $(\mu=-0.047)$; for dominance, it varied from -0.08 to -0.038 $(\mu=-0.058)$. These very small effect sizes (all less than one tenth of a standard deviation) indicate that variation in affect perception between paired speech and tEGG signals is practically indistinguishable from baseline variation in speech-affect perception. 

\section{Conclusion}\label{sec:conclusion}
In this paper, we proposed a novel signal processing method based on LPC and EGG to systematically remove phonetic content from a vocal signal while retaining as much of the signal as possible to preserve paralinguistic cues. Empirical comparison of perceived affect in response to original speech audio signals and corresponding transformed EGG  signals were shown to be highly similar, indicating that little affective information is lost as a result of our transformation method. As such, the algorithm we describe provides a method for generating a synchronized dataset of affective speech, EGG, and phoneme-agnostic audio signals that can be used to further research acoustic correlates of paralinguistic vocal cues.

Finally, we note that an important limitation on the generalizability of the method described here is the necessity of having parallel speech audio and EGG recordings, the latter being relatively rare. For this reason, we are currently using a dataset of speech, EGG, and transformed EGG stimuli (only part of which has been described here) toward training an artificial neural network model to isolate paralinguistic cues from speech audio without the need for EGG.

\section{Acknowledgements}
The authors thank Dr. Michael Frank, Dr. Perry Cook and Dr. Julius Smith for their support, guidance and feedback.

\bibliographystyle{IEEEbib}
\bibliography{mybib,extrabib}

\end{document}